%% file: main.tex
\documentclass[hidelinks,conference]{IEEEtran}
\pdfoutput=1
\usepackage{textcomp}
\usepackage{xcolor}
\usepackage{footnotebackref}
\usepackage{footmisc} 
\usepackage{mathtools}
\usepackage{multicol}
\usepackage{cuted}
\usepackage{lipsum, color}
\usepackage{tabularx,booktabs}
\usepackage{hhline}
\usepackage{colortbl}
\usepackage{color}
\usepackage{amssymb}  

\usepackage{balance}            
\usepackage{bigstrut}
\usepackage{cite}               
\usepackage{threeparttable}     
\newcolumntype{M}[1]{>{\centering\arraybackslash}m{#1}}  
\input{AKmacros.tex}

\IEEEoverridecommandlockouts        

\begin{document}

\bstctlcite{IEEEexample:BSTcontrol}

\title{The BRAM is the Limit: Shattering Myths, Shaping Standards, and Building Scalable PIM Accelerators}

\input{authorsIEEE}


\input{arxive-notice}

\maketitle

\input{abstract}
\input{intro}

\input{background}

\input{standard}

\input{casestudy}

\input{results}

\input{conclusion}

\bibliographystyle{IEEEtran}
\IEEEtriggeratref{5}
\IEEEtriggercmd{\balance}
\bibliography{IEEEabrv,ref}

\end{document}

%% file: AKmacros.tex
\usepackage{color} 
\newcommand{\Ax}{$\times$}  

\usepackage{pgffor}

\newcommand{\gemvName}{IMAGine}

\newcommand{\gemvElabUl}{\underline{I}n-\underline{M}emory \underline{A}ccelerated \underline{G}EMV eng\underline{ine}}

%% file: authorsIEEE.tex

{
\author{

\IEEEauthorblockN{
    MD Arafat Kabir\IEEEauthorrefmark{1}, 
    Tendayi Kamucheka\IEEEauthorrefmark{1}, 
    Nathaniel Fredricks\IEEEauthorrefmark{1},\\
    Joel Mandebi\IEEEauthorrefmark{2},
    Jason Bakos\IEEEauthorrefmark{3},
    Miaoqing Huang\IEEEauthorrefmark{1}, and
    David Andrews\IEEEauthorrefmark{1}
}
\IEEEauthorblockA{
\IEEEauthorrefmark{1}%
    Department of Electrical Engineering and Computer Science,
    University of Arkansas, \\ 
\IEEEauthorrefmark{3}%
    Department of Computer Science and Engineering,
    University of South Carolina, \\
\IEEEauthorrefmark{2}%
    Advanced Micro Devices, Inc. (AMD)
}

\{makabir, tfkamuch, njfredri, mqhuang, dandrews\}@uark.edu, 
jmandebi@amd.com, jbakos@cse.sc.edu,%
\thanks{This material is based upon work supported by the National Science Foundation under Grant No. 1955820.}
}   
}   

%% file: arxive-notice.tex

\onecolumn
{\large\vspace*{\fill}

© 2024 IEEE.  Personal use of this material is permitted.  
Permission from IEEE must be obtained for all other uses, 
in any current or future media, including reprinting/republishing this 
material for advertising or promotional purposes, 
creating new collective works, for resale or redistribution to servers or lists, 
or reuse of any copyrighted component of this work in other works. \\

This work is an extended version of a poster, presented at the 
2024 32nd IEEE International Symposium on Field-Programmable Custom Computing Machines (FCCM)
and will appear in the proceedings and on the IEEE website soon.

\vspace*{\fill}
}
\twocolumn

%% file: abstract.tex
\begin{abstract}
Many recent FPGA based Processor-in-Memory (PIM) architectures have appeared with promises of impressive levels of parallelism but with performance that falls short of expectations due to reduced maximum clock frequencies, an inability to scale processing elements up to the maximum BRAM capacity, and minimal hardware support for large reduction operations. 
In this paper, we first establish what we believe should be a ``Gold Standard" set of design objectives for PIM-based FPGA designs.  
This Gold Standard was established to serve as an absolute metric for comparing PIMs developed on different technology nodes and vendor families as well as an aspirational goal for designers.

We then present {\gemvName}, an {\gemvElabUl} used as a case study to show the Gold Standard can be realized in practice. 
{\gemvName} serves as an existence proof that dispels several myths surrounding what is normally 
accepted as clocking and scaling FPGA performance limitations.
Specifically, {\gemvName} clocks at the maximum frequency of the BRAM and scales to  
100\% of the available BRAMs.  
Comparative analyses are presented showing execution speeds over existing PIM-based GEMV engines on FPGAs and
achieving a \mbox{2.65\Ax\ -- 3.2\Ax} faster clock.
An AMD Alveo U55 implementation is presented that achieves a system clock speed of 737 MHz, 
providing 64K bit-serial multiply-accumulate (MAC) units for GEMV operation.
This establishes {\gemvName} as the fastest PIM-based GEMV overlay, outperforming even the custom PIM-based FPGA accelerators reported to date.
Additionally, it surpasses TPU v1-v2 and Alibaba Hanguang 800 in clock speed while offering an equal or greater number of MAC units.
\end{abstract}


%% file: intro.tex
\section{Introduction}
The exponential growth of network-connected devices (IoT) and social media applications 
has significantly changed the landscape of computing workloads.
Modern workloads, such as scientific computation, graph processing, and machine learning, generate and process datasets that
are expanding at a rate that outpaces Moore's Law\cite{tpuv4_2021}.
However, today's processors remain constrained by the ``Memory Wall'' of the von Neumann architecture,
which limits the ability to exploit the parallelism within these memory-intensive tasks. 
Processing-in-memory (PIM) or Compute-in-Memory (CIM) 
architectures are being pursued~\cite{zpim2020,pimca2023,sramcim2022,skhynixPim_2023,ccb2021,spar2021,comefa2022,comefa2023,bramac2023,m4bram2023,picasoposter2023,picaso2023}
to mitigate the memory wall and enable processing performance to scale with memory capacity.

Modern Field Programmable Gate Arrays (FPGAs) with 100s of Mbits of SRAM distributed throughout the device 
in the form of disaggregated memory resources can provide several TB/s of internal bandwidth.  This is an ideal programmable substrate for creating customized Processor In/Near Memory accelerators. 
Several PIM array-based accelerator designs~\cite{ccb2021,spar2021,comefa2022,comefa2023,bramac2023,m4bram2023}
have been proposed to harness this massive internal bandwidth.
However, results reported to date reinforce a long-held community belief that competitive overlays and not toy examples cannot clock at the maximum frequency of a BRAM or reach a compute density sufficient to compete with their custom 
Application Specific Integrated Circuit (ASIC) counterparts.  
These limitations have served as motivation for researchers to propose redesigned Block-RAM (BRAMs)-LUT integrated PIM tiles to increase compute densities of FPGAs.

While compute density is increased the maximum BRAM clock frequency is reduced similar to overlays.  Additional scalability issues are also introduced at the system level: a faster and larger device does not imply a faster system speed or a linear increase in the compute units with increased BRAM density.
While each proposal offers some relative improvement over prior designs,
there does not yet exist an absolute metric or yardstick that could be used to evaluate the efficiency of these designs. This has made it difficult to perform quantitative comparisons between PIM arrays implemented in different logic families from one vendor or between different vendors.

This paper first lays out a set of design objectives that form a ``Gold Standard'' or theoretical upper limit 
for BRAM-LUT-based PIM array architectures.  
This standard can serve as an absolute metric for comparing the efficiency of existing designs while also serving 
as an aspirational set of design objectives for future designs.   
The key implementation challenges for each objective are discussed and the approaches we employed to create {\gemvName}, 
a PIM design that approaches the theoretical performance of the Gold Standard.
Run time results show that {\gemvName} shatters some of the myths concerning performance limitations of FPGA overlays.
Our contributions can be summarized as follows,
\begin{itemize}
    \item A set of Gold Standard design goals for PIM array-based accelerators.
          We argue these goals need to be met to claim a ``Scalable High-Performance PIM design'' on FPGAs.
    \item We develop a case study that explores the design challenges and techniques that must be mastered to achieve those goals.
    \item We present {\gemvName}, an {\gemvElabUl}, as the fastest FPGA PIM-based GEMV accelerator that clocks faster than Google's TPU v1-v2.
    \item We study {\gemvName} and existing PIM-based FPGA accelerators to demonstrate how the proposed standard can be used as a guide to make near-optimal design choices.
\end{itemize}

Our designs will be published as open-source implementations and freely available 
for study, use, modification, and distribution without restriction.

%% file: background.tex
\section {Related Work}
\label{sec:background}

\subsection{Custom-BRAM PIMs}
Wang et al~\cite{ccb2021} proposed the Compute-Capable BRAM (CCB) as a PIM tile (or block) based on Neural Cache~\cite{neuralcache2018}. 
CCB exposes compute parallelism within a BRAM by converting each BRAM bitline into a bit-serial Processing Element (PE).
This comes at the cost of increased implementation complexity, as activating multiple wordlines 
requiring extra voltage supply and reduced voltage levels for robustness. 
CCB was used to build RIMA~\cite{ccb2021} to accelerate recurrent neural networks (RNNs).
RIMA achieved 1.25\Ax\ and 3\Ax\ higher performance compared to 
the Brainwave DL soft processor\cite{msbrainwaveNPU_2018} for 8-bit integer and block floating-point precisions, respectively. 
On a CCB-enhanced Stratix 10 FPGA, RIMA achieved an order-of-magnitude higher performance over a comparable GPU.

Arora et al~\cite{comefa2022,comefa2023} proposed CoMeFa.
Similar to CCB, CoMeFa employs bit-serial PEs per SRAM bitline but exploits 
the dual-port nature of a BRAM to read two operands without requiring voltage supply modifications.  Still, adjustments to Sense Amplifiers (SAs), additional latches for time multiplexing, and SA cycling are required.  To evaluate the performance and energy benefits of CoMeFa RAMs, various microbenchmarks, including 
General Matrix-Vector Multiplication (GEMV) and General Matrix-Matrix Multiplication (GEMM) were studied in~\cite{comefa2023}. Augmenting an Intel Arria 10-like FPGA with CoMeFa RAMs delivered a geomean speedup of 2.55× across diverse applications.

CCB and CoMeFa use a transposed layout~\cite{ccb2021,comefa2023} to store data along the bitlines in a column-major format,
introducing additional latency due to conversion from the original data's row-major format.
Chen et al addressed this issue 
in BRAMAC~\cite{bramac2023} and M4BRAM~\cite{m4bram2023}.
They avoid computing on the slow and power-intensive primary BRAM array by copying 
operands to a smaller ``dummy array" for MAC operations. 
BRAMAC requires 2-/4-/8-bit predefined weights and activations,
limiting its use to quantized uniform-precision deep neural nets. 
M4BRAM overcomes some of these limitations by enabling variable activation
precision between 2 and 8 bits with linearly scaled MAC latency.
BRAMAC was integrated into Intel's Deep Learning Accelerator (DLA)~\cite{intelDLA_2017} 
targetting Alexnet and ResNet-34 convolution models.
Combining BRAMAC-2SA/BRAMAC-1DA with Intel's DLA resulted in an average speedup 
of 2.05×/1.7× for AlexNet and 1.33×/1.52× for ResNet-34. 
Similarly evaluated using Intel's DLA, M4BRAM showed an average speedup of 2.16× with less than 0.5\%
accuracy loss using mixed-precisions compared to a full-precision baseline.
M4BRAM surpassed BRAMAC by an average of 1.43× across diverse benchmarks.

\subsection{BRAM-Overlay PIMs}

The proposed custom BRAM-based PIM architectures show promise for the future but are currently unavailable in mainstream FPGAs. 
To leverage the benefits of PIM architectures in contemporary FPGAs, PIM overlay architectures have been proposed.
Panahi et al~\cite{spar2020, spar2021, spar2thesis_2022} proposed a PIM overlay as part of the SPAR-2 accelerator,
connecting bit-serial PEs from the programmable fabric with BRAMs.
SPAR-2 PIM blocks contain a 4×4 PE array with North-East-West-South (NEWS) network support,
forming a 2D array of SIMD bit-serial processors at the system level.

SPAR-2 was implemented on Virtex-7 and Virtex UltraScale FPGAs with 10K PEs to accelerate
MLP, LSTM, GRU, and CNN~\cite{spar2thesis_2022} deep learning applications.
It achieved up to 34.2× and 3.5× speedups compared to other custom HLS-based and RTL-based accelerators, respectively.

Building upon the PIM overlay of SPAR-2 (SPAR2-PIM), Kabir et al proposed PiCaSO~\cite{picaso2023}
with configurable pipeline stages along the datapath. 
PiCaSO introduced an intermediate muxing module to enable zero-copy in-block reduction
and a ``binary-hopping'' pipelined NEWS network for array-level reduction.
PiCaSO provided competitive performance and memory utilization 
efficiency compared to both CCB and CoMeFa custom-BRAM architectures.

\subsection{PIM Emulators}
Mosanu et al~\cite{pimulator2022} proposed a synthesizable SystemVerilog model for
prototyping PIM architectures on their emulation platform PiMulator.
Their model is compatible with other FPGA-SoC frameworks, 
providing a high degree of usability in the FPGA and RISC-V ecosystems.

Dervay et al~\cite{cimulator2022} proposed a synthesizable RTL model of CIM (PIM) 
for their open-source emulation framework CIMulator.
They defined an instruction set for the model to integrate it with a RISC processor.
A fully functional processor/CIM system was demonstrated featuring 
cycle-accurate simulation/emulation, instruction-level energy estimation, 
and end-to-end program execution with guaranteed correctness.

%% file: standard.tex
\section{Defining A Gold Standard}
\label{sec:standard}
Table~\ref{tab:pim-speeds} summarizes the maximum achieved frequencies of the PIM designs discussed in section \ref{sec:background}.
From the relative frequency columns (Rel.), it is observable that the clock frequency \textsl{f$_{PIM}$} of all the PIM Tiles (Blocks), 
both overlays and custom BRAMs, are significantly slower compared to the maximum frequency for the device BRAMs (\textsl{f$_{BRAM}$}),
with the exception of PiCaSO.
Their system frequencies (\textsl{f$_{Sys}$}) are 2.1\Ax\ -- 3.7\Ax\ slower than the BRAM maximum frequencies (\textsl{f$_{BRAM}$}).
This decrease in system frequency was attributed to the limitations of the soft logic and the routing resources of the FPGAs.
It was also reported as unlikely that an FPGA accelerator
at the system level would operate at a frequency surpassing the degraded frequency (\textsl{f$_{PIM}$}) of these PIM designs,
even in a more advanced node than the evaluation platforms~\cite{comefa2022,comefa2023,m4bram2023,bramac2023}. 

Further observation yielded that most of these systems could not utilize all available BRAMs as PIMs.
This lower utilization combined with a lower clock frequency results in less efficient use of 
the available internal BRAM bandwidth of the devices and a lower system-level compute density. 
A final observation shows a troubling common pattern: as the utilization of BRAMs
increases the achievable system-level clock frequency decreases~\cite{ccb2021,comefa2023}.

These observations motivated our interest in understanding if these results were a new reality 
of BRAM PIM arrays or symptomatic of specific design decisions and implementation choices. 
This led us to ask two questions.  
What is the fastest frequency a PIM-based design should/could be able to achieve on FPGAs,
and was it possible to scale compute density up to the maximum BRAM capacity without degrading the clock frequency?

We posited that compute efficiency should be measured relative to an FPGA's full internal bandwidth 
and only limited by the density of the devices BRAMs. 
This would require a PIM architecture to run at the BRAM maximum speed with compute density 
scaling linearly up to the full density of BRAMs on a device. 
We termed such an ideal PIM architecture as our Gold Standard.

\input{pim-design-speeds}

\subsection{Ideal Clocking}
\label{sec:idealClock}
In FPGAs, BRAMs are the single component with the longest latency~\cite{v7datasheet_2021, ultrascale_datasheet_2021, stratix10_datasheet}, 
thus representing the timing bottleneck for setting clock speed.
Though in a typical FPGA design, the logic and routing delays can dominate the overall path delays,
at the unit level, FPGA resources like LUTs, FF, and routing blocks are much faster than the BRAM.
Thus, we define the maximum frequency (Fmax) of the BRAM as the Gold Standard clock frequency for the PIM accelerator.  
This requires processing elements along bit lines to be designed such that they do not degrade this frequency.

To assess the practicality of this standard, we closely examined two AMD FPGA families: Virtex-7 and UltraScale+. 
While Virtex-7 CLB resource's delay numbers are available in the datasheet~\cite{v7datasheet_2021},
those numbers are not publicly available for UltraScale+ devices.
Thus, we created a design where all timing paths are one logic level deep and averaged all paths to obtain Table~\ref{tab:celldelay}.
The Total column sums the cell delays in the columns to its left.
The BRAM column lists the clock period for BRAM Fmax. 
The Min column displays the minimum delay of a net passing through a switchbox. 
Net Budget is derived by subtracting the Total column from the BRAM column.
Comparing the net budget with the minimum net delay shows the possibility of designing 
at least two LUTs deep logic paths that can run at the BRAM Fmax on these device families.

In certain FPGA families, achieving this constraint may be challenging due to the presence of multiple dies 
and fixed-function blocks supporting various functionalities, such as
PLLs, high-performance IO blocks, PCIe blocks~\cite{ultrascale_lib_2021}, and application processors~\cite{zynqdatasheet_2022}.
These blocks impact placement and routing, eventually affecting the maximum achievable clock frequency.
In such devices, this Gold Standard may not be achievable.
However, FPGA vendors can employ this clocking standard to ensure the achievability of this constraint in device families targeting PIM designs.

\input{cell-delay}

\subsubsection{Design Challenges}

Achieving the Gold Standard for clock speed involves addressing challenges at the architectural level. 
System logic residing in the same clock domain as the BRAM should match or exceed the BRAM Fmax.
This may require adding optional pipelines in the design, which can be enabled
at a later stage of implementation if the logic depth is limiting the clock.
PIM array-based designs typically share control logic across multiple PIM blocks\cite{ccb2021,spar2021,spar2thesis_2022}, 
resulting in high fanout nets which can impact the system clock.
To address this issue, the control logic can be replicated, thereby reducing the fanout.
If high fanout is unavoidable, pipelined fanout trees should be synthesized for these signals.

Despite the aforementioned design considerations, timing failures may still occur during placement and routing due to long routes causing excessive net delays. 
To mitigate this, the system architecture should employ a tile-based approach
at the system level~\cite{ccb2021,comefa2022,comefa2023,spar2021} to localize logic and routing.
The RTL design should be implemented with an awareness of potential placement and routing issues.
If feasible, alternative implementations should be incorporated to be selected during the implementation stage to address placement and routing issues.

\subsection{Ideal Scaling of Peak-Performance}
The key advantage of a PIM architecture is the massive parallelism it can offer:
all concurrent bitlines of the memory array can be designed into concurrent processing elements~\cite{neuralcache2018, ccb2021, comefa2022, comefa2023}.
As a Gold Standard, we posited that the peak-performance of a PIM design needs to scale linearly with the on-chip BRAM count. 
Existing PIM designs do not adhere to this standard.

\begin{figure}
\centering
\includegraphics[width=\linewidth]{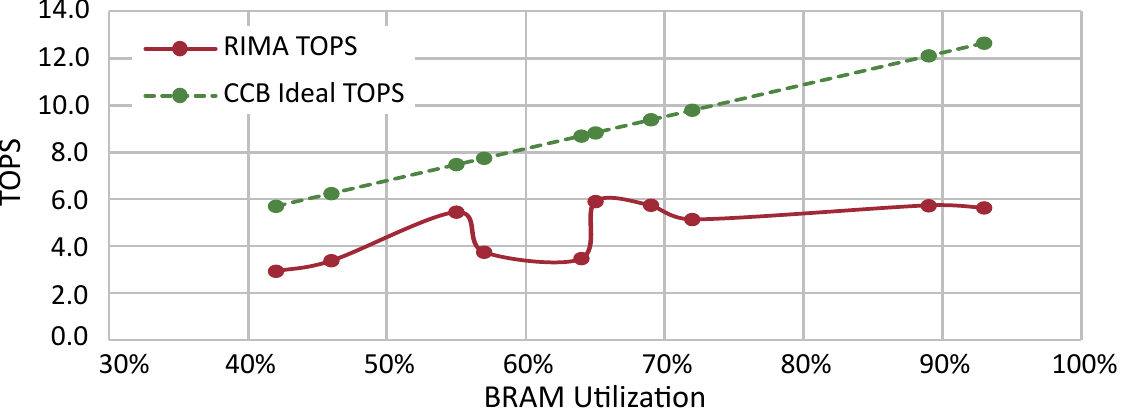}
\caption{Ideal scaling vs. actual TOPS of RIMA on Stratix 10 GX2800}
\label{fig:stdscaling}
\end{figure}

The compute capacity in custom-BRAM-based PIM designs~\cite{ccb2021, comefa2022, comefa2023, bramac2023, m4bram2023} 
scales linearly with BRAM count if all BRAM tiles are used in PIM mode.
However, a significant sacrifice is imposed in the clock frequency that ends up limiting the achievable peak-performance on the device. 
Table~\ref{tab:pim-speeds}  \textsl{f$_{PIM}$} column indicates that the custom-BRAM PIM designs run up to 2.5× slower than the BRAM Fmax.
In addition, their system frequency decreases with an increase in BRAM usage.
Fig.~\ref{fig:stdscaling} plots RIMA's peak-performance from Table-II of~\cite{ccb2021}, 
computed using reported BRAM utilization and M-DPE clock frequency. 
The irregular trend is attributed to RIMA's system-level architecture.
If RIMA adhered to the ideal scaling Gold Standard, even at the degraded CCB frequency of 624~MHz,
its peak-performance would align with the CCB Ideal TOPS line. 
The gap between these plots represents wasted compute capacity and memory bandwidth provided by CCB BRAMs.

\subsubsection{Design Challenges}

In PIM accelerators, PIM blocks handle computation, while the rest of the system primarily 
manages data pipelines and control logic~\cite{ccb2021,spar2021,comefa2022,comefa2023}.
The main obstacle to ideal scaling in existing PIM accelerators is the high logic utilization in the rest of the system.
The same design principle as in ideal clocking must be applied here as well: the BRAM should be the utilization bottleneck, not the control logic.
To prevent the control logic from becoming the utilization bottleneck, controllers should be shared between multiple PIM blocks.
Control signals should be designed to minimize the number of unique control sets in the system.
A control set is the group of control signals (set/reset, clock enable, and clock) for a register or latch~\cite{ctrlset_guideline}.
Excessive unique control sets can degrade placement, impacting system scalability and clock frequency. 
Overall, the rest of the system should complement the PIM array without limiting its scalability and performance.

\subsection{Ideal Reduction Latency}
\begin{figure}
\centering
\includegraphics[width=\linewidth]{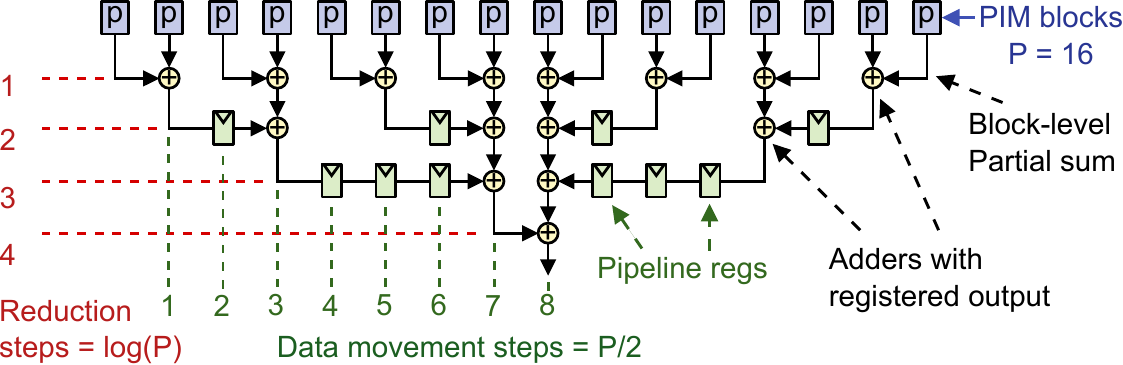}
\caption{Reduction latency breakdown of a pipelined binary adder tree maintaining ideal clocking constraint.}
\label{fig:accumtree}
\end{figure}

Reduction is a critical operation in applications like GEMM and deep learning as it requires data movement throughout the distributed BRAM memories and/or processing elements of the PIM array.
Accumulation, a common reduction operation, is often implemented using a pipelined
adder tree~\cite{ccb2021, comefa2023} as shown in Fig.~\ref{fig:accumtree}.
Adder trees are resource-hungry, especially on routing, requiring more adders and routing resources as the PIM row size increases.
Bit-parallel implementations are even harder to manage compared to bit-serial: INT8 requires 8× more 
logic and routing resources than bit-serial implementations. 
Such high utilization can subsequently affect system frequency and scalability.

To prevent the reduction network from becoming the utilization and routing bottleneck, 
sharing logic and routing resources with the rest of the system may be necessary.
Existing high-performance reduction architectures for FPGA 
implementation~\cite{ling_designing_2005, zhuo_highperformance_2007, tai_accelerating_2012, huang_modular_2013, tang_resource_2021, ullah_highperformance_2022} 
highlight the unavoidable trade-off between reduction latency and resource utilization.
To guide this trade-off, we propose the Gold Standard for reduction latency as the following empirical model,
\begin{align}
\label{eq:goldaccum}
\text{Array-Level Reduction}_{\text{gold}} &= aN \log P + bP + c \\
\label{eq:goldaccumblk}
\text{In-Block Reduction}_{\text{gold}} &= aN \log k 
\end{align}
In this context, a PIM ``block'' is a single BRAM tile, like CCB, BRAMAC, PiCaSO, etc.,
along with its related logic.
In-Block reduction generates partial sums accumulating the PEs in a PIM block, while array-level reduction accumulates these partial sums.
Here, $P$ = no. of partial sums obtained from all PIM columns involved in the reduction process, $N$ = operand width (precision),
$k$ = no. of PE columns in a PIM block;
$a, b, c$ are implementation-specific parameters with their ideal range specified in Table~\ref{tab:redGoldParam}.

\input{reduction-gold-param}

The intuition behind~\eqref{eq:goldaccum} and the parameter ranges is explained using Fig.~\ref{fig:accumtree}.
The total reduction (accumulation) latency can be broken down into two parts: reduction operation (add) and data movement.
The term $aN \log P$ represents the latency of reduction operations (add) only, requiring at least $\log P$ reduction steps;
the base of the $\log$ represents the number of operands reduced per step, typically 2.
The PE architecture (bit-serial, bit-sliced, or bit-parallel) determines the value \mbox{of $a$}. 
The lower bound of $a$ is 1/N because at least one cycle is needed per reduction step ($aN \geq 1$). 
We set the upper bound for $a$ to 2 because bit-serial PEs, especially in overlays, 
commonly require 2 cycles to process each bit ($aN = 2N$) of the operand~\cite{spar2021, spar2thesis_2022, picaso2023}.

Under the ideal clocking constraint, data-movement latency depends linearly 
on the number of PIM columns in a row, represented \mbox{by $bP$}.
Assume each PIM block (p) in Fig.~\ref{fig:accumtree} occupies a large enough area
such that a net does extend beyond two consecutive PIM blocks without violating the ideal clocking constraint.
Then constructing the adder tree without sacrificing the clock speed requires
one pipeline stage to be placed along each PIM column as in Fig.~\ref{fig:accumtree}.
This requires P/2 cycles ($b = 1/2$) for accumulation towards the middle in this example.
In general, the value of $b$ depends on the speed of FPGA's routing resource relative to BRAM Fmax.
Typically $b<1$ in modern FPGAs with fast routing resources allowing nets to span multiple PIM columns at BRAM Fmax.
The lower bound of $b$ is 0, corresponding to a bit-parallel pipelined reduction tree that perfectly
overlaps data movement with computation.

In \eqref{eq:goldaccum}, $c$ represents delays outside the reduction network.  
The In-Block partial-sum generation latency~\eqref{eq:goldaccumblk} is a simpler version \mbox{of \eqref{eq:goldaccum}},
where $k$ denotes the number of PE columns accumulated within the PIM block.
A term for pipelined data movement is absent as signals can move within a block without violating the ideal timing constraint. 
In-Block reduction latency \eqref{eq:goldaccumblk} can be absorbed into $c$,
\mbox{making \eqref{eq:goldaccum}} as the overall reduction latency standard.
The lower bound of $c$ is 0, representing no additional cycles outside of the reduction network.
Having no upper bound allows the architecture to vary without violating the Gold Standard;
some preprocessing steps may be employed such as the pop-count-based adder in RIMA~\cite{ccb2021},
before entering the reduction network that requires tens or hundreds of cycles.

Though the proposed standard \eqref{eq:goldaccum} closely resembles a pipelined adder tree,
it can provide insights and help identify design inefficiencies in various other architectures. 
Importantly it can steer designers towards optimal or near-optimal reduction network designs. 
We explore this in Section~\ref{sec:results} through a quantitative study of existing designs.

\input{reduction-latency-eqn}

\subsubsection{Design Approaches}
Table~\ref{tab:redlatency} shows the latency equations for bit-serial PIM architectures~\cite{ccb2021,spar2021,comefa2023,picaso2023}
and their relative utilization of resources. 
The bit-parallel architectures~\cite{bramac2023,m4bram2023} are not listed because the block-level
reduction is built into their MAC units and they do not have a proposed implementation for array-level reduction.

SPAR-2~\cite{spar2thesis_2022} implements two reduction approaches: 
linear shift through a NEWS network of PE columns followed by either linear-add or binary-add.
The binary-add approach was an optimization over the linear-add to reduce the number of add operations.
Block-level accumulation in~\cite{ccb2021,comefa2023} involves across-bitline copies, bit-serial add, and a pop-count-based adder.
Its $\log^2$(k) term is due to the increase in the precision by 1-bit after each iteration.
Based on the limited discussion in~\cite{ccb2021}, we assume a perfect ``global reduction tree''
with 2 cycles pipeline overhead, and has the array-level latency presented in CCB/CoMeFa row in Table~\ref{tab:redlatency}.
The binary-hopping reduction network in~\cite{picaso2023} utilizes a NEWS network with register stages between PIM blocks, 
enabling direct copying between distant blocks without intermediate writes;
a pipelined muxing module eliminates the overhead of across-bitline copies in block-level reduction.

The NEWS network is the slowest but simplest, requiring minimal resources, 
while the global reduction tree is the fastest but demands the highest resources. 
Binary-hopping NEWS achieves faster speed than vanilla NEWS at the expense of higher utilization. 
These implementations inherently trade off latency and resource utilization, 
aligning with the proposed standard~\eqref{eq:goldaccum}. 
In the analysis section, we will quantify these latencies
and show how the standard can help identify design inefficiencies and guide toward optimal design.

%% file: pim-design-speeds.tex
\begin{table}
\setlength{\tabcolsep}{2.0pt}
\caption{Maximum Frequency (MHz) of Existing FPGA-PIM Designs}
\label{tab:pim-speeds}
\centering
\begin{tabular}{c|ccc|cc|cc}
\hline
PIM Design & Type   & Device & \textsl{f$_{BRAM}$}   & \textsl{f$_{PIM}$}    & Rel.   & \textsl{f$_{Sys}$}  & Rel. \bigstrut\\
\hline
CCB        & Custom  & Stratix 10  & 1000   & 624    & 62\%   & 455    & 46\% \bigstrut[t]\\
CoMeFa-A   & Custom  & Arria 10    & 730    & 294    & 40\%   & 288    & 39\%   \\
CoMeFa-D   & Custom  & Arria 10    & 730    & 588    & 81\%   & 292    & 40\%   \\
BRAMAC-2SA & Custom  & Arria 10    & 730    & 586    & 80\%   & -      & -      \\
BRAMAC-1DA & Custom  & Arria 10    & 730    & 500    & 68\%   & -      & -      \\
M4BRAM     & Custom  & Arria 10    & 730    & 553    & 76\%   & -      & -      \\
SPAR-2     & Overlay & UltraScale+ & 737    & 445    & 60\%   & 200    & 27\%   \\
PiMulator  & Overlay & UltraScale+ & 737    & -      & -      & 333    & 45\%   \\
PiCaSO     & Overlay & UltraScale+ & 737    & 737    & 100\%  & -      & -      \\
\hline
\end{tabular}%
\end{table}

%% file: cell-delay.tex
\begin{table}
\centering
\setlength{\tabcolsep}{4.0pt}
\caption{Delay (ns) Breakdown of 1-level Logic Path in AMD Devices}
\label{tab:celldelay}
\begin{threeparttable}
\begin{tabular}{cccc|c|ccc}
\hline
       & FF-C2Q\tnote{1} & LUT\tnote{2} & FF-Setup & Total\tnote{3}  & BRAM\tnote{4} & Net Budget & Min\tnote{5} \bigstrut\\
\hline
V7     & 0.290   & 0.34   & 0.255  & 0.885  & 1.839   & 0.954 & 0.272 \bigstrut[t]\\
US+    & 0.087   & 0.15   & 0.098  & 0.335  & 1.356   & 1.021 & 0.102 \bigstrut[b]\\
\hline
\end{tabular}%
\begin{tablenotes}
\item[1] Clock-to-Q delay of flip-flops
\item[2] Average of delay through LUTs 
\item[3] Total cell delay
\item[4] BRAM pulse-width requirement, clock period for Fmax
\item[5] Minimum net delay through a switchbox
\end{tablenotes}
\end{threeparttable}
\end{table}

%% file: reduction-gold-param.tex
\begin{table}
\caption{Parameters of Gold Standard for Reduction Latency}
\label{tab:redGoldParam}
\centering
\begin{tabular}{ccc}
\hline
Parameter & Ideal Range & Related to \bigstrut\\
\hline
$a$      & 1/N $\leq a \leq$ 2 & Latency of reduction steps (addition) \bigstrut[t]\\
$b$      & 0   $\leq b \leq$ 1 & Latency of data movement \\
$c$      & 0 $\leq$ c          & Cycles spent outside reduction network \bigstrut[b]\\
\hline
\end{tabular}%
\end{table}

%% file: reduction-latency-eqn.tex
\begin{table}
\centering
\renewcommand{\arraystretch}{1.2}
\setlength{\tabcolsep}{2.5pt}
\caption{Reduction/Accumulation Latency of Existing PIM Designs}
\label{tab:redlatency}
\begin{threeparttable}
\begin{tabular}{cccc}
\hline
       & Block Level & PIM Array Level & Utilization\tnote{1} \bigstrut\\
\hline
Linear-Add\tnote{2} \cite{spar2thesis_2022}  & 3N (k-1)                     & 3N (P-1)                   & Low \bigstrut[t]\\
Binary-Add\tnote{2} \cite{spar2thesis_2022}  & 2N$\log$(k) + N(k-1)         & 2N$\log$(P) + N(P-1)        & Low  \\
CCB/CoMeFa~\cite{ccb2021}                    & 2N$\log$(k) + $\log^2$(k)    & $\log$(P) + 2              & High \\
Binary-hopping \cite{picaso2023}             & (N+4) $\log$(k)              & (N+4) $\log$(P) + P-1       & Medium \\
\bfseries Proposed Standard                  &  $a$N$\log$(k)               & $a$N$\log$(P) + $b$P + $c$  & Balanced\tnote{3} \bigstrut[b]\\
\hline
\end{tabular}%
\begin{tablenotes}
\item[1] Qualitative logic and routing resource utilization
\item[2] NEWS network with linear shift then add
\item[3] Offers latency vs. resource utilization trade-off \\
\end{tablenotes}
\end{threeparttable}
\end{table}

%% file: casestudy.tex
\section{Gold Standard Case-Study: {\gemvName}}
\label{sec:casestudy}
While the Gold Standard is an ultimate objective to strive for, it can also serve as a framework 
for crafting an efficient architecture of a PIM accelerator on FPGAs.
In this section, we discuss the architecture of {\gemvName}, a PIM array-based GEMV accelerator,
that will be implemented and analyzed in Section~\ref{sec:results} on an off-the-shelf FPGA.
{\gemvName} serves a dual purpose: it provides a case study demonstrating the use of the Gold Standard
to make better design choices and evaluate how achievable those standards are in a practical design.

\begin{figure}
\centering
\includegraphics[width=\linewidth]{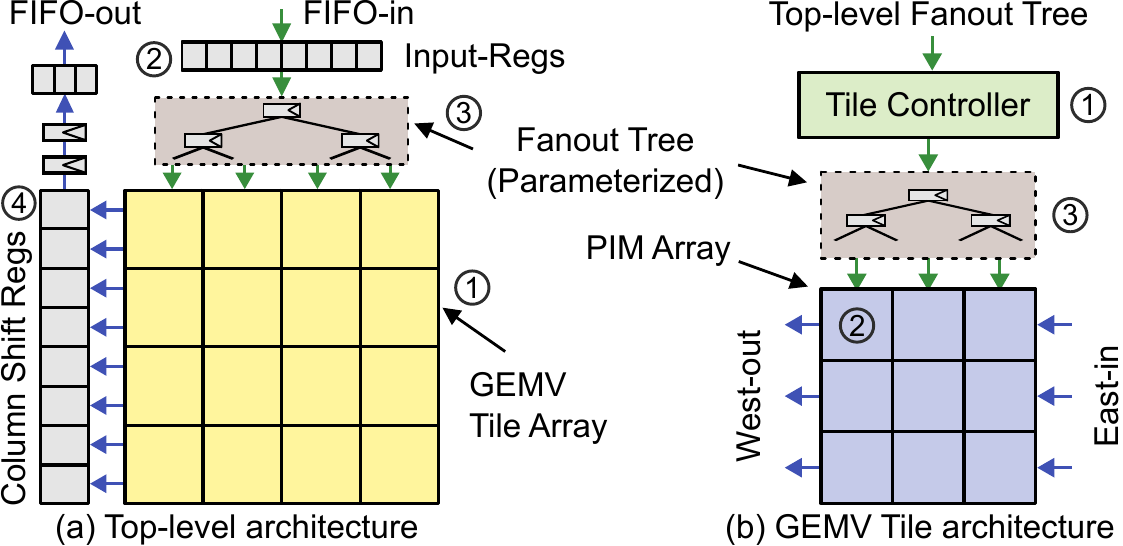}
\caption{
System architecture of {\gemvName} illustrating the data and instruction flow
(a) through the GEMV engine and
(b) within GEMV tiles.
}
\label{fig:archSysTile}
\end{figure}

\subsection{System-Level Architecture}
The top-level system is illustrated in Fig.~\ref{fig:archSysTile}(a).
It consists of (1) a 2D array of GEMV tiles, (2) a set of input registers, (3) a fanout tree connecting
the input registers to the tile array, and (4) a column of shift-registers to read out the final result.
The input registers are used by the front-end processor to send instructions to the GEMV tile controllers.
The fanout tree is parameterized to be adjusted during implementation.
The 2D tile array is implemented as a parameterized module that instantiates and connects GEMV tiles
to build the tile array. 
The bits written to the register file of the leftmost PE in the array are shifted into the column shift registers.
At the end of the GEMV operation, the output vector is stored in the column shift registers,
which can be shifted up and read through the FIFO-out port, one element per cycle.

\subsection{{\gemvName} Tile Architecture}
\label{sec:tilearch}
Illustrated in Fig.~\ref{fig:archSysTile}(b), the GEMV tile is the heart of {\gemvName}.
It consists of (1) an FSM-based controller, (2) a 2D array of PIM blocks, and (3) a fanout tree between them.
The controller receives the instruction written to the input registers at the top level,
decodes it, and generates the sequence of control signals needed to execute the instruction.
The fanout tree connects the control signals to all PEs in the PIM array
and is parameterized for adjustment during implementation. 
The PIM array interfaces allow cascading with arrays in neighboring tiles on each side. 
During accumulation, partial results move from east to west through PIM arrays, 
ultimately accumulating in the left-most PE column of the left-most GEMV tile in a row.

\subsection{Tile Controller}
\begin{figure}
\centering
\includegraphics[width=\linewidth]{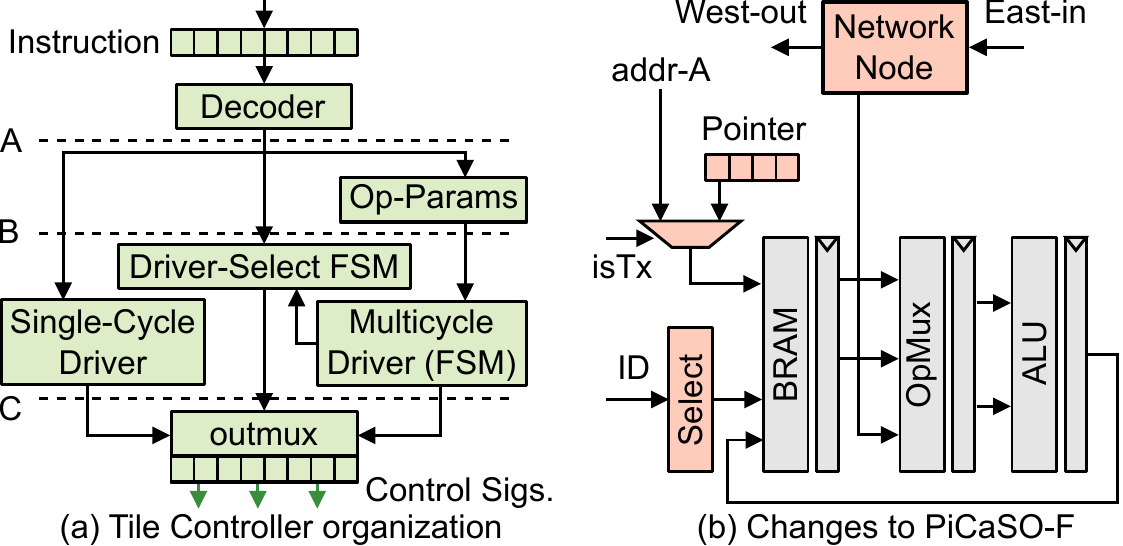}
\caption{
Architectures of 
(a) GEMV controller and 
(b) PiCaSO-IM, the adapted version of PiCaSO-F~\cite{picaso2023}.
}
\label{fig:archControlPim}
\end{figure}

Fig.~\ref{fig:archControlPim}(a) shows the architecture of the tile controller.
As discussed in Section~\ref{sec:idealClock}, logic paths need to be short enough to achieve the ideal clock rate.
However, estimating precise logic depth during RTL design is challenging and 
the requirement varies across devices.
Thus, we grouped the combinatorial logic into meaningful steps and added optional pipeline stages
illustrated by the dashed lines A, B, and C in Fig.~\ref{fig:archControlPim}(a).
Running synthesis, we ensured that each step could be implemented in one or two logic levels.

The controller takes a 30-bit instruction, which is executed by either the single-cycle or the multicycle driver.
The 2-state driver-selection FSM enables any one of them based on the opcode.
The single-cycle driver can execute one instruction every cycle, while the multicycle driver takes several
cycles to execute instructions like ADD, SUB, MULT, etc. 
including an additional cycle to load its parameters from the Op-Params module.
All inputs and outputs are registered to localize timing paths within the controller.

\subsection{PIM Module}
\label{sec:modpim}
We based the design of the {\gemvName} PIM module called PiCaSO-IM on PiCaSO~\cite{picaso2023} for the following three reasons: 
(1) it is publicly available and open-source~\cite{github_picaso}, 
(2) it is a modifiable overlay that can be ported and studied on existing AMD devices, and 
(3) PiCaSO-F, a pipelined configuration of PiCaSO, already clocks at the BRAM Fmax.

Modifications to PiCaSO-F were needed to enhance its control capabilities to implement the GEMV tile.
These modifications are highlighted in red in Fig.~\ref{fig:archControlPim}(b). 
The additional logic supporting the original NEWS network was removed,
keeping only the parts needed for east-to-west data movement.
PiCaSO-F lacked control signals for selectively enabling/disabling a block required in {\gemvName}.
Block-ID-based selection logic was included in PiCaSO-IM.
Our accumulation algorithm needed 3 addresses to maximize the overlap of data movement and computation.
As PiCaSO-F supported only 2 simultaneous addresses, we added a pointer register for the third address.

If PiCaSO is realized as a custom-BRAM tile as proposed in~\cite{picaso2023}, these changes can be
implemented in programmable logic fabric, keeping registerfile, OpMux, and ALU modules within the BRAM tile.
We name such a custom-BRAM implementation of PiCaSO-IM as PiCaSO-CB.

%% file: results.tex
\section{{\gemvName} Implementation and Analysis}
\label{sec:results}

In this section, we discuss the bottom-up implementation of {\gemvName},
setting the design goal to be the Gold Standard discussed in Section~\ref{sec:standard}.
In~\cite{picaso2023}, PiCaSO was studied on AMD Alveo U55C (xcu55c, -2 speed grade).
We use the same device as our implementation platform to keep the results predictable.
The BRAM Fmax on this device is 737~MHz~\cite{ultrascale_datasheet_2021}, which sets the target clock period to be 1.356~ns.
All of the following studies were carried out using Vivado 2022.2.

\input{picaso-mod}

\subsection{PiCaSO-IM Block}
We first verified that the additional logic added to the original PiCaSO-F did not degrade the 
BRAM Fmax within the PIM block or create a logic utilization bottleneck.
A 4\Ax4 array of the new PiCaSO-IM was tested and compared against
the numbers reported in~\cite{picaso2023}. This comparison is shown in Table~\ref{tab:picasomod}.
The Change\% row shows the utilization and clock speed change compared to the original implementation.
BRAM utilization is 0.5 because a PiCaSO block uses one RAMB18 tile,
which is reported by Vivado as 0.5 number of RAMB36 tile of AMD devices.
As observed, the modifications did not affect the clock frequency and the utilizations of BRAM and DSP.
The flip-flop utilization increased by only 10.6\%.
Though there is a significant increase (74.7\%) in the LUT utilization, the overall Slice utilization only increased by 8.6\%.
This means that the additional logic has a high packing factor in the logic slices.
As a result, the additional logic would not be expected to introduce a utilization bottleneck.

\subsection{{\gemvName} GEMV Tile}
Before we implemented the GEMV tile discussed in Section~\ref{sec:tilearch}, its
components were studied individually to verify if they met the design requirements.
The GEMV tile contains a 12\Ax2 PIM array and 2 stages of pipeline in the fanout tree 
This configuration best fits the physical layout of the Alveo U55 FPGA as discussed later in this section.
Table~\ref{tab:tile} shows the utilization and performance of these components and
their relative values compared to the entire GEMV tile.

The controller together with the fanout network passed the timing constraints at a clock rate up to 890 MHz.
Because the PIM array contains the BRAM, it cannot run faster than the BRAM Fmax.
It passed the timing at 737 MHz, which is the ideal clock for Alveo U55 according to the Gold Standard.
As observed in Table~\ref{tab:tile}, the logic utilization of the controller is around 5\% of 
the entire tile and requires no DSPs, while around 90\% of the logic resources are consumed by the PIM array.
Thus, the controller and the fanout tree are not expected to bottleneck system frequency or utilization.
The GEMV tile's speed and scalability are fundamentally dependant on the PIM array, which is the desired outcome.

\input{tile-util-perf}

\subsection{Scalability Study}
\begin{figure}
\centering
\includegraphics[width=\linewidth]{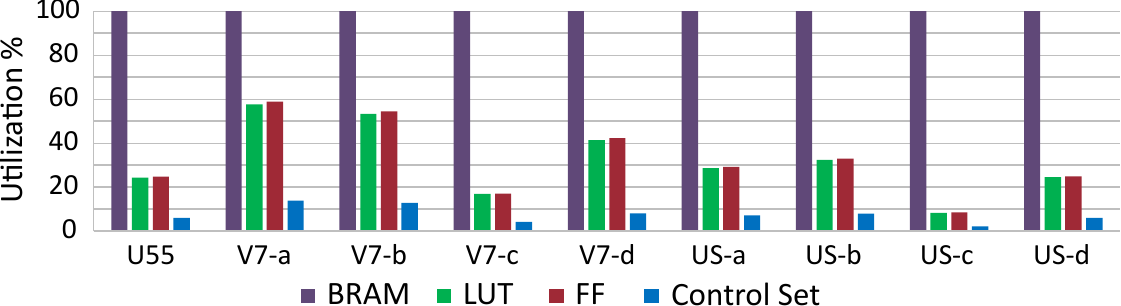}
\caption{Resource usage of {\gemvName} on representatives of Virtex-7 and Ultrascale+ families utilizing 100\% BRAMs as PIM overlays.}
\label{fig:deviceScale}
\end{figure}

To evaluate the scalability of our architecture on different device families, we followed the approach in~\cite{picaso2023}.
Along with Alveo U55, four representatives were selected from AMD's Virtex-7 and UltraScale+ devices based on two criteria: BRAM capacity and LUT-to-BRAM ratio.
Table~\ref{tab:devices} lists these devices with their BRAM capacity, LUT-to-BRAM ratio, and a short ID used in Fig.~\ref{fig:deviceScale}.
These are the same devices on which the scalability of PiCaSO-F was studied. 
The target clock frequency of the system was set to 100 MHz on all devices to avoid timing issues 
and only focus on the logic utilization of the system at this point.

Fig.~\ref{fig:deviceScale} shows a bar graph of post-implementation utilization numbers of {\gemvName} 
on the representative devices. 
As observed, {\gemvName} can utilize 100\% of the available BRAMs as PIM overlays
providing 64K PEs in U55, with only 25\% logic and 6\% control set utilization.
This leaves sufficient logic resources to implement the fanout trees and pipeline 
stages if they are needed to achieve the target clock speed.
In fact, {\gemvName} scaled up to 100\% of available BRAM in all the representative devices for Virtex-7 and UltraScale+ families.

In the Virtex-7 family, the device V7-a has the smallest number of BRAMs and the smallest LUT-to-BRAM ratio.
{\gemvName} used around 60\% logic resources to provide 24K PEs in V7-a.
In the UltraScale+ family, US-a and US-b have the smallest number of BRAMs and the smallest LUT-to-BRAM ratio, respectively.
In these devices {\gemvName} provide 23K and 67K PEs, respectively, using roughly 30\% logic resources.
For devices with more BRAMs and a higher LUT-to-BRAM ratio the logic utilization is very small:
the logic utilization in US-c is less than 10\% providing 69K PEs.

If we can keep this scaling and run the PIM blocks at BRAM Fmax, the Gold Standard of linear scaling of peak performance would be met.
Compared to the scalability study presented in~\cite{picaso2023}, the logic utilization of {\gemvName} is roughly 11\% more 
than just the PiCaSO array on average across all the devices.
This increase is due to the controller logic, data pipelines, and the additional logic implemented in PiCaSO-IM.

\input{device-scale-tab}

\subsection{System-Level Timing Optimizations}

For the final implementation, the target clock was set to the Gold Standard for Alveo U55
with a period of 1.356~ns to match the BRAM Fmax.
The goal of the study was to find out how close we can get to the Gold Standard, 
and what are the practical challenges that limit us from achieving the Gold Standard.
Achieving the best performance on a device always requires several iterations of device-specific optimizations.
In the first iteration, we started with GEMV tiles having a 4\Ax4 PIM array,
without the fanout tree between the controller and the array.
After going through the implementation flow with the default settings of Vivado and 
a few optimization iterations, we achieved a setup slack of -0.52~ns.
The critical paths were within the controller with a logic depth of 4, going through the pipeline stage A
of the controller as shown in Fig.~\ref{fig:archControlPim}(a).
So, we enabled the pipeline stage A in the controller and moved forward with the second iteration.

At the end of the second iteration of implementation, we achieved a setup slack of -0.38~ns.
The critical nets were the control signals between the controller and the PIM array.
These nets were failing the timing requirement because of high fanout and long routes 
of the control signals between the controller and the PIM array.
Thus, we synthesize a fanout tree between the controller and the PIM array
empirically choosing 2 levels and a fanout of 4 for the next iteration.

\begin{figure}
\centering
\includegraphics[width=\linewidth]{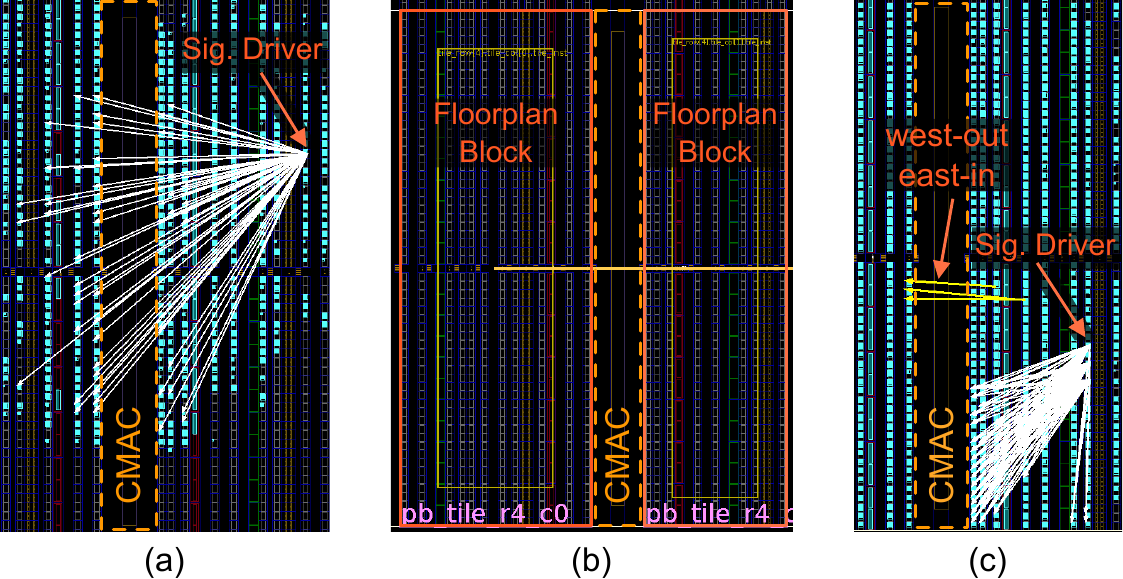}
\caption{
Avoiding unnecessary hard-block (CMAC) crossing
(a) placement and net connections before floorplanning,
(b) floorplanning to localize logic and routing,
(b) placement and net connections in the final design.
}
\label{fig:floorplan}
\end{figure}

The design achieved a setup slack of -0.27~ns in the third iteration.
This time, we had to take a closer look at the design to reveal the main reason for the timing failures.
Alveo U55 contains several hardened blocks including high-performance Ethernet port (CMAC)~\cite{alveoU55ug_2023}.
Most of the failing paths were due to the long routes crossing such hard blocks.
The white lines in Fig.~\ref{fig:floorplan}(a) highlight some of those critical nets crossing a CMAC block.
To avoid placement results creating such paths, we created floorplanning blocks for each tile 
to localize the placement of logic and routing within a region dedicated to the tile.
The floorplanning blocks were placed avoiding those hard blocks as shown in Fig.~\ref{fig:floorplan}(b).
This required defining a tile with the PIM array dimension of 12\Ax2 for Alveo U55.

Fig.~\ref{fig:floorplan}(c) shows the placement and net connections in the final iteration.
The logic and routing of each tile were localized on either side of the hard block;
the white lines representing the high fanout control signals do not cross the CMAC block. 
Only the logically essential nets, the inter-tile connections for east-to-west accumulation, 
cross the CMAC block requiring minimal routing resources.
The yellow lines in Fig.~\ref{fig:floorplan}(c) highlight some of those inter-tile nets.

All paths in the final design met the timing requirement at 737~MHz clock,
which demonstrates that the ideal clocking standard is practically achievable.
Utilizing 100\% available BRAMs as PIMs, this design also achieved the ideal linear scaling of peak-performance.
Surprisingly, this clock rate is faster than custom GEMM accelerator ASICs TPU v1-v2~\cite{tpuv4_2021} 
and Alibaba Hanguang 800~\cite{alibabaNPU2020}, that run at 700~MHz.
Both Alveo U55 and TPU v2 are manufactured at 16~nm and Hanguang 800 at 12nm technology nodes.
So, this clock improvement is not due to a technology node difference. 
On Alveo U55, {\gemvName} has an equal number of PEs compared to TPU v1 (64K), and 4\Ax\ of TPU v2 (16K).
However, {\gemvName} can only deliver up to 0.33 TOPS which is significantly smaller compared to 
TPU v1 (92 TOPS) and v2 (46 TOPS) due its bit-serial architecture. 
This makes a compelling case: even though an FPGA design will probably never outperform custom ASICs in terms of 
peak-performance or performance-per-watt, the right set of design goals and guiding principles can 
bring it very close in terms of clock speed and compute density.
Our proposed Gold Standard can serve that purpose for PIM array-based FPGA designs.

\begin{figure*}
\centering
\includegraphics[width=\linewidth]{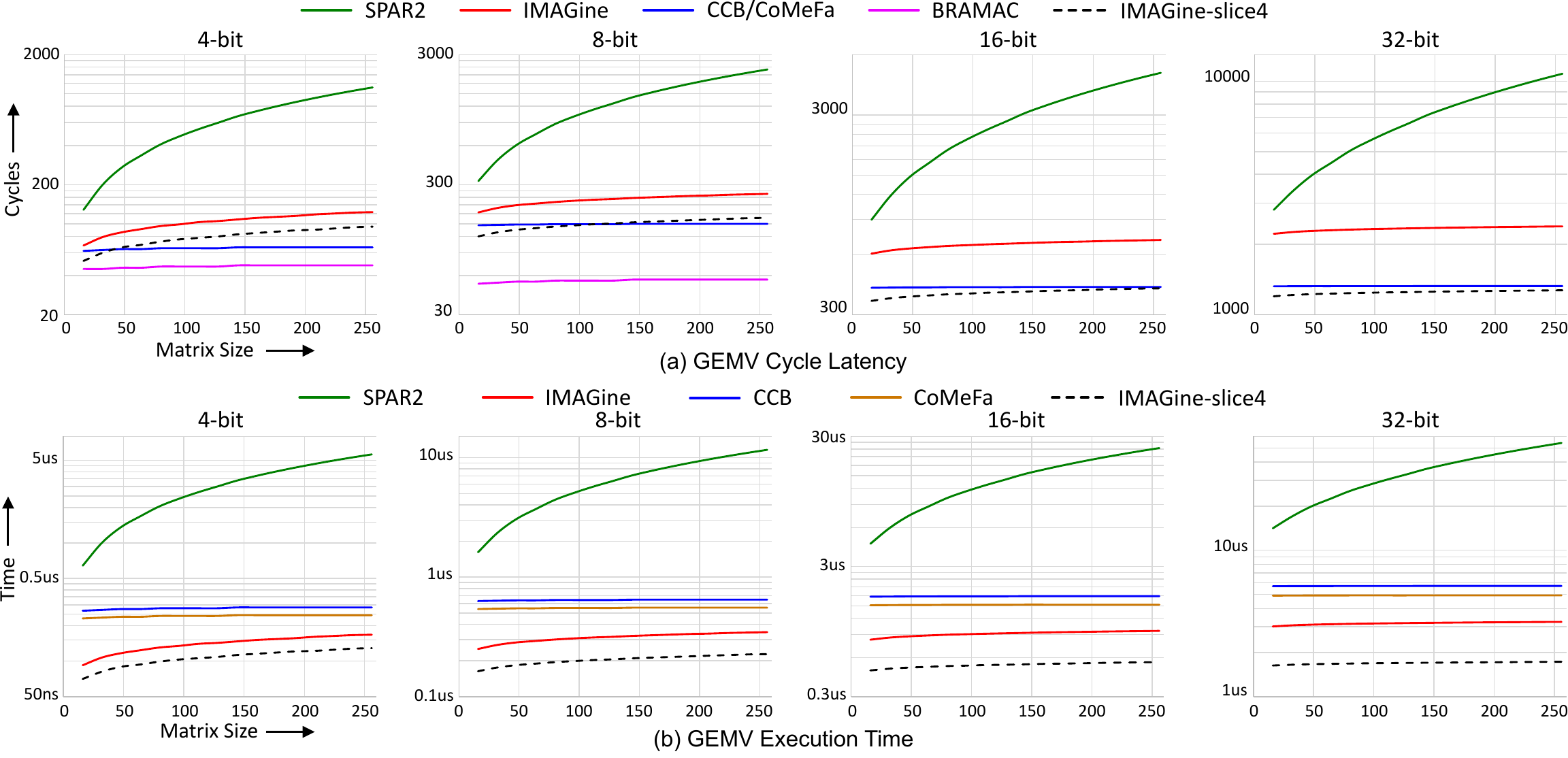}
\caption{Cycle latency and execution time of GEMV operation on different PIM array-based FPGA accelerators}
\label{fig:gemvlatency}
\end{figure*}

\input{system-comparison}

\subsection{Comparison With Other PIM Accelerators}
Table~\ref{tab:systemcomp} shows the utilization and system frequencies of existing GEMV engines and equivalent PIM-based systems.
System-level utilizations and frequencies for BRAMAC and M4BRAM-based systems were not reported in~\cite{bramac2023, m4bram2023}.
Though RIMA is specialized for accelerating RNNs, 
a major part of the system implements GEMV operation using Dot Product Engines (M-DPEs)~\cite{ccb2021}.
The RIMA numbers are taken from Table~II of~\cite{ccb2021} for comparison, 
which was evaluated on a Stratix 10 GX2800 FPGA with a BRAM Fmax of 1~GHz~\cite{stratix10_datasheet}.
Its fastest reported configuration (RIMA-Fast) runs at 455~MHz, which is 2.2\Ax\ slower than the BRAM Fmax. 
The largest reported configuration (RIMA-Large) utilizes 93\% of BRAMs and runs at 278~MHz,
4\Ax\ slower compared to BRAM Fmax. 
The GEMV/GEMM systems based on CCB and CoMeFa 
were evaluated on an Arria 10 GX900 with a BRAM Fmax of 730~MHz~\cite{comefa2023}.
Though CoMeFa-based designs run slightly faster than the CCB-GEMV engine,
they are still roughly 3\Ax\ slower than the BRAM Fmax of the device.
Thus, neither of the CCB and CoMeFa-based GEMV/GEMM engines scaled well at the system level.

SPAR-2~\cite{spar2021} utilized only 30\% of the BRAMs while running 4\Ax\ slower than BRAM Fmax on both tested platforms.
Thus, its performance and scalability are even worse than CCB and CoMeFa-based systems.
On the other hand, {\gemvName} has a system clock running at the BRAM Fmax while utilizing 100\% device BRAM as PIMs. 
Thus {\gemvName} takes advantage of the full internal bandwidth offered by the BRAMs in the device.
As a PIM-based GEMV engine, {\gemvName} not only outperformed all existing designs 
but also verified the Gold Standard in terms of BRAM Fmax clock frequency and peak-performance 
scalability up to the full internal bandwidth is achievable. 
This is an important proof of concept design that dispels earlier beliefs that 
overlays cannot achieve BRAM Fmax clock frequencies at the system level.
It is the fastest PIM-based GEMV engine implemented on any FPGA, 
running at a clock rate 2.65\Ax\ -- 3.2\Ax\ faster than any existing design.

As observed in Table~\ref{tab:systemcomp}, RIMA and CCB/CoMeFa-based GEMV engines 
exhaust either the logic resources or the DSPs of the device even though 
their PIM blocks are implemented by customizing the BRAM tile itself.
Even after being an overlay, {\gemvName} is achieving faster clock and better scalability using 0 DSPs and 
only one-third of the device logic resources due to its near-optimal architectural choices guided by the Gold Standard.
Like SPAR-2, {\gemvName} does not use DSPs to implement the bit-serial PEs.
With a custom-BRAM PIM module like PiCaSO-CB discussed in Section~\ref{sec:modpim}, 
{\gemvName} would consume about 10\% of device resources while being fully scalable and implementable even in resource-limited FPGAs.

\subsection{GEMV Execution Latency}
Fig.~\ref{fig:gemvlatency}(a) displays GEMV cycle latency for the PIM designs showing 
matrix dimensions (matrices are square) on the x-axis and cycle latency in log scale on the y-axis.
The execution times shown in Fig.~\ref{fig:gemvlatency}(b) are computed by multiplying cycle latencies with the corresponding clock periods
of CCB GEMV, CoMeFa-D GEMM, SPAR-2 (US+), and {\gemvName} from Table~\ref{tab:systemcomp} system frequencies.
We adopted the approach in~\cite{bramac2023} to model the block-level cycle latencies of
CCB, CoMeFa, BRAMAC, and SPAR-2 using their analytical models.
{\gemvName}'s latency model was developed and validated through cycle-accurate simulations of the system.
The global reduction tree of RIMA was modeled as an adder tree, perfectly pipelined with the last iteration of in-block accumulation.
For CoMeFa and BRAMAC GEMV latency, we assumed the same system as RIMA but with CoMeFa or BRAMAC replacing the CCB blocks.
For SPAR-2, only the latency for the binary-add version is shown because 
the linear-add version is too slow to plot together with the other systems.

As observed in Fig.~\ref{fig:gemvlatency}(a), BRAMAC has the shortest cycle latency,
due to their hybrid bit-serial \& bit-parallel MAC2 algorithm.
MAC latency in BRAMAC grows linearly with operand bit-width,
while it grows quadratically in bit-serial architectures like CCB, CoMeFa, SPAR-2, and PiCaSO.
This result is not surprising; domain-specific architectures can deliver the best performance at a very low cost.
BRAMAC supports only 2, 4, and 8-bit precisions particularly targetting low-precision 
applications like quantized neural network acceleration.
However, BRAMAC is less suitable for general computing tasks like GEMV for full-precision scientific computing 
or even neural networks requiring wider precisions.
BRAMAC~\cite{bramac2023} did not report their system-level frequency which is why we could not plot its execution time.

In all precisions, SPAR-2 has the longest cycle latency and execution time due to its slow NEWS accumulation network.
Its accumulation latency increases almost linearly with the matrix dimension.
CCB and CoMeFa-based GEMV engines have the shortest cycle latency among bit-serial architectures across all precisions.
This is due to their fast reduction algorithm based on a pop-count adder and pipelined adder tree.
The cycle latency of {\gemvName} is significantly shorter compared to SPAR-2 but longer than CCB/CoMeFa-based implementations.
However, {\gemvName} has the fastest clock, which is at a minimum 2\Ax\ faster than any of the other GEMV engines. 
When accounting for the clock frequency, {\gemvName} outperforms all other GEMV engines in terms of overall execution time.
This highlights the importance of the system clock speed over the cycle latency.
Although the reduction tree-based approach in the CCB/CoMeFa GEMV engines has the shortest cycle latency, 
the slower clock ultimately degrades the end-to-end latency times below that of {\gemvName}.

\subsection{Further Improvements}

While {\gemvName} outperforms the existing designs, this does not guarantee it is the optimal implementation. There could still be some room for improvement, particularly in the reduction network.
When examining all of the architectures, we can intuitively understand which designs are better suited for the reduction process.
However, the proposed Gold Standard can quantify these insights even without the architectural knowledge of the implementations.
Since the studied systems employ different reduction approaches, we define the reduction latency for the GEMV operation as 
any cycle spent outside the multiplication stage, encompassing the In-Block latency~\eqref{eq:goldaccumblk}.
We curve-fitted the proposed model~\eqref{eq:goldaccum} to the reduction cycle latencies of the designs in Fig.~\ref{fig:gemvlatency}(a).
Table~\ref{tab:goldfit} shows the fitted parameters for 32-bit operands and their interpretations.

\input{gold-fit}

SPAR-2 linear-add parameters significantly deviate from their ideal ranges outlined in Table~\ref{tab:redGoldParam},
revealing a suboptimal reduction network design.
SPAR-2 binary-add has $a$ in the standard range, which means the reduction operation is at least near optimal.
This was achieved through an optimization reported in~\cite{spar2thesis_2022} to reduce the number of add operations.
A high value of $b$ in both approaches indicates that accumulation latency is dominated by the data movement part,
which is notably slower than the ideal.
The value of $c$ being 0 indicates the reduction process does not involve extra cycles
outside the reduction network, which is true by design.

For CCB/CoMeFa, both parameters are near the smallest possible values: here 0.3 $\approxeq$ 1/N, N=32.
This implies it has the shortest possible cycle latency.
The value of $c$ approximates the In-Block accumulation and pipeline setup latency of 202 cycles
spent generating the 16 partial sums per block~\cite{ccb2021} before they enter the reduction network.

{\gemvName}'s parameters fall within the standard range, implying at least a near-optimal implementation.
In this case, as well, the value of $c$ approximates the In-Block accumulation latency of 144 cycles.
Since $b$ is near its upper bound, it indicates room for improvement in the data movement part.
Because {\gemvName} is utilizing only 30\% of the logic resources in U55, we can modify the network
to implement 2-bit or 4-bit bit-sliced accumulation, possibly without affecting the system clock speed.
This has the potential to further improve the cycle latency of the GEMV operation.
Additionally, the PEs can be modified to implement Booth's Radix-4 instead of the default Radix-2 algorithm adopted from PiCaSO.

The {\gemvName}-slice4 curves in Fig.~\ref{fig:gemvlatency} shows the GEMV latency of a variant of {\gemvName} 
with a 4-bit sliced accumulation network and a PE implementing Booth's radix-4 multiplication.
This latency is estimated by adjusting the analytical model of {\gemvName} assuming no effect on the clock rate.
In terms of cycle latency, it can run almost as fast as CCB/CoMeFa-based GEMV implementations.
Because of the higher system frequency, it will then significantly outperform all other state-of-the-art PIM-based GEMV accelerators.
This example demonstrates how the proposed Gold Standard can aid in identifying 
design inefficiencies and making optimal design choices for PIM-based accelerators in FPGAs.

%% file: picaso-mod.tex
\begin{table}
\setlength{\tabcolsep}{3.5pt}
\caption{Utilization and Clock Frequency of Modified PiCaSO 4$\times$4 Tile}
\label{tab:picasomod}
\centering
\begin{tabular}{ccccccc}
\hline
       & LUT    & FF     & Slice  & DSP    & BRAM   & Max Freq. \bigstrut\\
\hline
PiCaSO-F Tile   & 774    & 1799   & 243    & 0      & 8      & 737 MHz \bigstrut[t]\\
PiCaSO-F Block  & 49     & 113    & 15     & 0      & 0.5    & 737 MHz \\
\hline
PiCaSO-IM Tile  & 1352   & 1989   & 264    & 0      & 8      & 737 MHz \bigstrut[t]\\
PiCaSO-IM Block & 85     & 125    & 18     & 0      & 0.5    & 737 MHz \\
\hline
Change \% & 74.7\% & 10.6\% & 8.6\%  & -      & 0.0\%  & 0\% \bigstrut[t] \\
\hline
\end{tabular}%
\end{table}

%% file: tile-util-perf.tex
\begin{table}
\setlength{\tabcolsep}{2.0pt}
\caption{Utilization and Frequency of 12$\times$2 GEMV Tile Components}
\label{tab:tile}
\centering
\begin{tabular}{c|cc|cc|cc|c}
\hline
       & Controller &  Rel. & Fanout &  Rel. & PIM Array &  Rel. & Tile \bigstrut\\
\hline
LUT    & 167    & 5.8\%  & 0      & 0.0\%  & 2736   & 94.2\% & 2903 \bigstrut[t]\\
FF     & 155    & 4.0\%  & 615    & 15.9\% & 3096   & 80.1\% & 3866 \\
DSP    & 0      & -      & 0      & -      & 0      & -      & 0 \\
BRAM   & 0      & 0.0\%  & 0      & 0.0\%  & 12.0   & 100.0\% & 12 \\
Freq. (MHz) & 890   & 1.2$\times$   & 890    & 1.2$\times$   & 737    & 1$\times$     & 737  \bigstrut[b]\\
\hline
\end{tabular}%
\end{table}

%% file: device-scale-tab.tex
\begin{table}
\centering
\begin{threeparttable}
\caption{Representatives of Virtex-7 and Ultrascale+ Families~\cite{picaso2023}}
\label{tab:devices}
\begin{tabular}{cccccc}
\hline
Device & Tech   & BRAM\# & Ratio\tnote{1} & Max PE\#\tnote{2}  & ID \bigstrut[t] \\
\hline
xcu55c-fsvh-2   & US+    & 2016   & 646    & 64K      & U55 \bigstrut[t]\\
xc7vx330tffg-2  & V7     & 750    & 272    & 24K      & V7-a \\
xc7vx485tffg-2  & V7     & 1030   & 295    & 32K      & V7-b \\
xc7v2000tfhg-2  & V7     & 1292   & 946    & 41K      & V7-c \\
xc7vx1140tflg-2 & V7     & 1880   & 379    & 60K      & V7-d \\
xcvu3p-ffvc-3   & US+    & 720    & 547    & 23K      & US-a \\
xcvu23p-vsva-3  & US+    & 2112   & 488    & 67K      & US-b \\
xcvu19p-fsvb-2  & US+    & 2160   & 1892   & 69K      & US-c \\
xcvu29p-figd-3  & US+    & 2688   & 643    & 86K      & US-d \\
\hline
\end{tabular}%
\begin{tablenotes}
\item[1] LUT-to-BRAM ratio
\item[2] Number of PEs utilizing all BRAMs as PIMs
\end{tablenotes}
\end{threeparttable}
\end{table}

%% file: system-comparison.tex
\begin{table}
\setlength{\tabcolsep}{3.5pt}
\caption{Utilization and Frequency of PIM-based GEMV/GEMM Engines}
\label{tab:systemcomp}
\centering
\begin{threeparttable}
\begin{tabular}{c|cc|cc|cc}
\hline
       & LUT    & FF     & DSP    & BRAM   & \textsl{f$_{Sys}$}\tnote{1} & Rel. Freq  \bigstrut\\
\hline
RIMA-Fast & \multicolumn{2}{c|}{60\%} & 50\%   & 55\%   & 455    & 45.5\% \bigstrut[t]\\
RIMA-Large & \multicolumn{2}{c|}{89\%} & 50\%   & 93\%   & 278    & 27.8\% \\
CCB GEMV & \multicolumn{2}{c|}{27.9\%} & 90.1\% & 91.8\% & 231    & 31.6\% \\
CoMeFa-A GEMV & \multicolumn{2}{c|}{27.9\%} & 90.1\% & 91.8\% & 242    & 33.2\% \\
CoMeFa-D GEMM & \multicolumn{2}{c|}{25.5\%} & 92.4\% & 86.7\% & 267    & 36.6\% \\
SPAR-2 (US+) & 11.3\% & 2.4\%  & 0.0\%  & 14.5\% & 200    & 27.1\% \\
SPAR-2 (V7) & 28.5\% & 7.0\%  & 0.0\%  & 30.4\% & 130    & 23.9\% \\
{\gemvName} & 35.6\% & 24.8\% & 0.0\%  & 100.0\% & 737    & 100.0\% \\
{\gemvName}-CB\tnote{2} & 10.1\% & 7.2\%  & 0.0\%  & 100.0\% & 737    & 100.0\% \bigstrut[b]\\
\hline
\end{tabular}%
\begin{tablenotes}
\item[1] System frequency in MHz
\item[2] {\gemvName} with custom-BRAM PiCaSO-F (PiCaSO-CB)
\end{tablenotes}
\end{threeparttable}
\end{table}

%% file: gold-fit.tex
\begin{table}
\setlength{\tabcolsep}{3.5pt}
\caption{Curve-Fitted Parameters of Eqn.~\eqref{eq:goldaccum} for 32-bit Accumulation}
\label{tab:goldfit}
\centering
\begin{tabular}{c|ccc|cc}
\hline
          & \multicolumn{3}{c|}{Fitted Value} & \multicolumn{2}{c}{Speed Interpretation} \bigstrut[t]\\
                   & a      & b      & c      & Addition (a) & Movement (b) \bigstrut[b]\\
\hline
SPAR-2 Linear-Add  & 0      & 96     & 0      & Very Slow & Very Slow \bigstrut[t]\\
SPAR-2 Binary-Add  & 2      & 32     & 0      & Standard & Very Slow \\
CCB/CoMeFa         & 0.03   & 0.02   & 203.1  & Fast   & Fast \\
IMAGine            & 1.2    & 0.9    & 143    & Standard & Standard \bigstrut[b]\\
\hline
\end{tabular}%
\end{table}

%% file: conclusion.tex
\section{Conclusions and Future Work}
\label{sec:conclusion}
Processor In/Close to Memory PIM/CIM architectures have become popular frameworks 
replacing classic von Neumann architectures within domain-specific machine learning accelerators.  
PIM overlays as well as new reconfigurable fabrics for next-generation FPGAs are being 
explored with results showing projected clock frequencies much lower than the maximum 
for a BRAM and continuing to degrade as the number of PIM blocks scales.  
Furthermore, most current designs do not provide efficient hardware support between PIM blocks 
for the important reduction steps within GEMV matrix-vector operations.

This paper presented a study that established a theoretical upper limit
for PIM array-based architectures in BRAM-LUT-based FPGAs as an aspirational Gold Standard. 
It was then shown how the Gold Standard can be used for comparing and evaluating the efficiency of existing designs, 
both overlays and next-generation PIM reconfigurable fabrics. 
Important design challenges and implementation techniques were explored that can 
affect a specific design's ability to reach the performance of the Gold Standard.
Results were presented verifying the Gold Standard is achievable in FPGA-based overlays as well as redesigned BRAM-LUT PIM tiles. 
The {\gemvName} case study showed the full bandwidth of the BRAMs could be exploited and operated at BRAM's maximum frequency.

Scalability studies were presented, demonstrating that processing capacity scales linearly 
with increasing BRAM density, even for devices with low LUT-to-BRAM ratios.
{\gemvName}, a PIM array-base GEMV accelerator, with 64K PEs was implemented and run on Alveo U55, which achieved clock speed faster 
than the Tensor Processing Unit (TPU v1-v2) and Alibaba Hanguang 800.  
This breaks the myth that FPGA overlays and fabrics must clock slower than ASIC designs.

Although {\gemvName} achieved better performance compared to state-of-the-art FPGA-based PIM architectures 
and clocks faster than some ASIC accelerators, there are additional improvements that require additional exploration. 
Further improvement in end-to-end GEMV latency can be achieved by overlapping data broadcasting and 
multiply-accumulate operations, which will benefit deep learning applications.
We are currently working on a new MLIR-based compiler framework for hardware/software codesign 
and application-specific customization of {\gemvName}-like PIM array-based accelerators.